\documentclass[runningheads]{cl2emult}

\usepackage{makeidx}  
\usepackage{graphicx} 
\usepackage{subeqnar} 
\usepackage{multicol} 
\usepackage{cropmark} 
\usepackage{eso}      
\makeindex            

\begin{document}
\title*{The VIRMOS-VLT Deep Survey}
\toctitle{The VIRMOS-VLT Deep Survey}
%
%
\titlerunning{The VIRMOS-VLT Deep Survey}
%
\author{O. Le F\`evre\inst{1}
\and G. Vettolani\inst{2}
\and D. Maccagni\inst{3}
\and D. Mancini\inst{4}
\and A. Mazure\inst{1}
\and Y. Mellier\inst{5}
\and J.P. Picat\inst{6}
\and M. Arnaboldi\inst{4}
\and S. Bardelli\inst{9}
\and E. Bertin\inst{5}
\and G. Busarello\inst{4}
\and A. Cappi\inst{9}
\and S. Charlot\inst{7}
\and G. Chincarini\inst{8}
\and S. Colombi\inst{5}
\and B. Garilli\inst{3}
\and L. Guzzo\inst{8}
\and A. Iovino\inst{8}
\and V. Le Brun\inst{1}
\and M. Longhetti\inst{8}
\and G. Mathez\inst{6}
\and P. Merluzzi\inst{4}
\and H.J. McCracken\inst{1}
\and R. Pell\`o\inst{6}
\and L. Pozzetti\inst{9}
\and M. Radovich\inst{5}
\and V. Ripepi\inst{5}
\and P. Saracco\inst{8}
\and R. Scaramella\inst{9}
\and M. Scodeggio\inst{3}
\and L. Tresse\inst{1}
\and G. Zamorani\inst{9}
\and E. Zucca\inst{9}
}
\authorrunning{Olivier Le F\`evre et al.}
%
%
\institute{Institut d'Astrophysique de Marseille, Marseille, France
\and IRA-CNR, Bologna, Italy
\and ICFTR-CNR, Milan, Italy
\and Osservatorio Astronomico di Capodimonte, Naples, Italy
\and Institut d'Astrophysique de Paris, Paris, France
\and Observatoire Midi-Pyr\'en\'ees, Toulouse, France
\and Max Planck fur Astrophysik, Garching, Germany
\and Osservatorio Astronomico di Brera, Milan, Italy
\and Osservatorio Astronomico di Roma, Rome, Italy
\and Osservatorio Astronomico di Bologna, Bologna, Italy}

\maketitle              

\begin{abstract}
The aim of the VIRMOS VLT Deep Survey (VVDS) is to study of the
evolution of galaxies, large scale structures and AGNs from a 
sample of more than 150,000 galaxies with measured redshifts in the range 
$0\leq z \leq 5+$. 
The VVDS will rely on the VIMOS and NIRMOS wide field 
multi-object spectrographs, which the VIRMOS consortium is delivering
to ESO. Together, they offer unprecedented multiplex capability in 
the wavelength range $0.37-1.8\mu$m, allowing for large surveys to
be carried out. The VVDS has several main aspects:
(1) a deep multi-color imaging survey over $18$deg$^2$ of more than
one million galaxies,
(2) a "wide" spectroscopic survey with more than 130,000 redshifts measured for objects
brighter than $I_{AB}=22.5$  over $18$deg$^2$, 
(3) a "deep" survey with 50,000 redshifts measured to $I_{AB}=24$, (4)
"ultra-deep" surveys with several thousand redshifts measured to
$I_{AB}=25$,
(5) multi-wavelength observations with the VLA and XMM.
\end{abstract}

\section{Introduction}

The statistical study of the properties of galaxies, active galactic
nuclei, or large scale structures, and their evolution, is a key to 
progress in our understanding of the evolution of the universe. 
In the past decade, our exploration of the universe has reached 
redshifts up to $z\sim1.2$
(\cite{lilly1},\cite{olf1},\cite{col1},\cite{ell1}), then up to redshifts $3-4$ (\cite{stei1}). 
Together, these data have given the first picture of the
star formation history in the universe (\cite{mad1}).

However, available samples remain small. Only a few thousand galaxies have
securely measured redshifts beyond $z\sim0.5$ where most of the
evolution appears to have taken place. While this provides a
first evaluation of the state of the galaxy population at these
redshifts, the current samples are too small to provide the statistical basis
required to investigate the dependence of evolution versus
key parameters such as the local galaxy/matter density, the luminosity/mass  
of galaxies, or galaxy types.
To trace galaxy evolution in more significant detail, it is necessary to 
conduct deep redshift surveys of  large
samples at increasingly large look-back times.
 
We describe here the VIRMOS VLT Deep  Survey (VVDS)
which is aiming to gather more than $1.5\times10^5$ 
redshifts for galaxies with $0<z<5+$. This project is
based on the Visible and
the Near-IR Multi-Object spectrographs (VIMOS and NIRMOS) being
built for the ESO-Very Large Telescope by the VIRMOS consortium. 
This survey will allow us to study evolution
over 90\% of the current age of the universe
with unprecedented details. 

\section{Survey goals}
The goal of the VVDS is to study the evolution of galaxies,
AGNs, and large scale structures, over a redshift range
$0<z<5+$. The requirement is to be
able to analyse the basic properties such as the luminosity function or
spatial correlation function of the galaxy population, as a function
of galaxy type or local density, in each of several time steps covering
the above redshift range. The number of galaxies necessary to make
a detailed study of the population  of galaxies can be estimated e.g. from
the computation of the luminosity function: considering that 50 galaxies are necessary
to measure the number of galaxies per magnitude bin, in 10 magnitude bins,
for 3 basic colors (type) and 3 types of environments from low to high density,
in 4 fields to beat cosmic variance, and in 7 time steps, the total number of 
galaxies required is thus $50\times10\times3\times3\times4\times7=126,000$
galaxies.

\section{The VIMOS and NIRMOS instruments}
The VIRMOS consortium of French and Italian institutes
led by the Institut d'Astrophysique in Marseille is building
two wide field multi-object spectrographs for the ESO-VLT, in
partnership with the following institutes: IRA-CNR and OABo in Bologna, 
IFCTR and OABr in Milan, OMP in Toulouse, Haute Provence Observatory, OAC in Naples.

VIMOS covers the wavelength range 0.37--1 microns, while NIRMOS
covers  1--1.8 microns. These imaging spectrographs will allow one 
to obtain wide field images, multi-slit spectroscopy, or integral field
spectroscopy data \cite{olf2}. 
Both instruments will make 
use of slit masks, custom-cut with a high power laser
machine capable to cut 200
slits in 15 minutes \cite{conti}. 

The main emphasis of these 2 instruments is the ability to observe large 
numbers of spectra simultaneously. The optimised optical layout in 4 separate
channels allows one to gather more than 800 spectra at once with VIMOS,
or close to 200 with NIRMOS, an unprecedented multiplex gain. Spectral resolutions
from $R=200$ to $R=2500$ will be available on VIMOS (one arcsecond slits, 
higher resolutions are possible with narrower slits), 
while NIRMOS will operate only
at resolutions higher than $R=2500$ to be able to go in between the
bright night sky emission lines. The field of view of VIMOS 
is $4\times(7\times8)$
arcmin$^2$, while the NIRMOS spectroscopic field is $4\times(6\times8)$ arcmin$^2$. 
VIMOS is in final integration phase at Observatoire de Haute Provence
(fig.1). It will undergo
commissioning at the Paranal observatory in the spring of 2001, and is being offered to the ESO
community as a facility instrument on VLT telescope \#3 starting in July 2001.

\section{Wide field deep imaging survey}
A deep imaging
survey in UBVRIKs complete to a depth $I_{AB}=25$  ($5\sigma$
detection for extended sources in a 3 arcsec aperture) 
is currently underway to prepare the unbiased photometric
sample of more than 1 million galaxies
from which spectroscopic targets will be selected. 

A total of 18 deg$^2$ in five fields is being covered,
4 fields $2\times2$ deg$^2$ each selected at high galactic latitude
around the celestial equator, and a $1\times2$ deg$^2$ field
around the Chandra deep field south (CDFS, \cite{giac}), see the list
at http:$//www.astrsp-mrs.fr/virmos/virmos\_deep\_survey.htm$). 
Each field allows to probe
scales up to $\sim100$h$^{-1}$Mpc. The 0226-04 field has been
selected to coincide with the high visibility area of XMM, 
and it is adjacent to the NOAO deep imaging survey
area \cite{jan}. This field is also subject to deep VLA observations
with  a $5\sigma$ sensitivity of the order of
$90\mu$Jy. 
In each field except the CDFS, we are acquiring UBVRI data, as well as
K' data for a smaller 900 arcmin$^2$ area. We are acquiring only I band
data on the CDFS at this time. 

The depth is set  to probe the luminosity function
of galaxies to M*+3 at z=1 and to select lyman-break and QSO candidates
out to redshifts $z>3-5$.
Particular care has been given to set a limiting magnitude
deep enough to allow the selection of magnitude limited samples for the spectroscopic
survey, free of bias and systematics. The limiting magnitude
of $I_{AB}\leq25$ at $5\sigma$ in a 3 arcsec aperture indeed represents
the magnitude at which 95\% of all galaxies with this magnitude
are detected and measured. Therefore, at the limit $I_{AB}=24$ of the deep
spectroscopic survey, all galaxies including those with low
surface brightness should be detected, and 
no bias is expected when conducting critical measurements like the
luminosity function or star formation rate. 

The 0226-04 field is observed at 
a depth 0.6 magnitudes deeper than the other fields as it 
will be the subject of the "deep" and "ultra-deep"
spectroscopic surveys.

See \cite{olf3} for more details on  the imaging survey.

\section{The VIRMOS spectroscopic survey}
We are planning to observe the following 
magnitude limited samples:\\
1. "Wide" survey: more than 
$10^5$ galaxies with redshifts measured to $I_{AB}=22.5$ (redshift up to
$z\sim1.3$), in 5 fields  \\
2. "Deep" survey: more than $4 \times 10^4$ galaxies with redshifts 
to $I_{AB}=24$ (redshift
up to $z\sim5$) in the 0226-04 field.\\
3. ``Ultra deep survey'': more than
$10^3$ galaxies with redshifts to $I_{AB}=25$  in the 0226-04 and
Hubble Deep Field South\\

The spectroscopic samples will be observed with $R\sim250$ in a first pass.
A subsample of 10,000 galaxies will then be selected for spectroscopy at a resolution
$R\sim2500-5000$  to study the evolution of the fundamental plane,
and perform a detailed analysis of the spectro-photometric properties 
of galaxies.

Subsamples of galaxies with specific color selection criteria
will also be selected, including Lyman-break galaxy candidates, QSOs
and extremely red objects (EROs). The logical flow of the
survey is presented in Fig.2.

\section{Preparing for data processing and analysis}
A data processing pipeline is being developed to 
process the multi-spectra data. It automatically
corrects for instrument signature, removes sky signal, 
extracts and calibrates 1D spectra in flux and wavelength.

A database has been built under Oracle8 
to handle the imaging and spectroscopic
catalog data, the images and spectra, as well as to cross correlate the
survey data to other surveys. 

The VVDS science team is developing the simulation and software tools 
necessary to analyse the survey in an efficient manner. We are integrating
all analysis tools into a common "toolbox" environment accessible through
a web interface connected to the database.

Strong emphasis 
is being put in quality control of the various processing
and analysis packages to ensure bug-free and optimal performances. 

\section{Summary}
The VIRMOS VLT Deep Survey is a major program to study the
evolution of galaxies, large scale structures, AGNs, clusters
of galaxies over more than 90\% of the age of the universe.
The VIMOS and NIRMOS instruments will provide unprecedented
capabilities. The VVDS survey is expected to officially start
in the summer of 2001, and will ultimately deliver more than 
150,000 redshifts of galaxies and AGNs in the distant universe.

\begin{figure}
\centering
\includegraphics[width=0.56\textwidth]{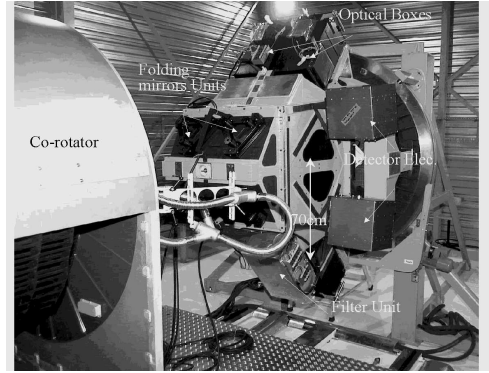}
\caption[]{The Visible Multi-Object Spectrograph (VIMOS) in final integration phase}
\label{eps2}
\end{figure}

\begin{figure}
\centering
\includegraphics[width=0.72\textwidth]{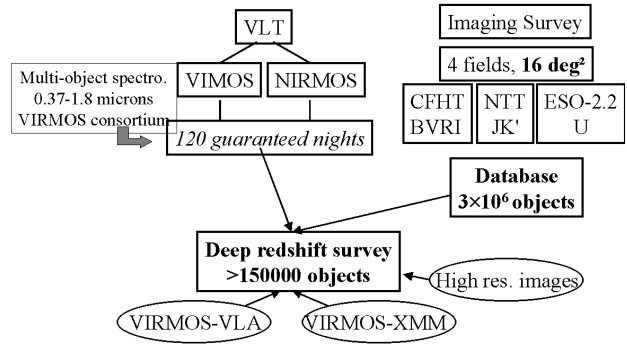}
\caption[]{VIRMOS-VLT Deep Redshift Survey: logical flow}
\label{eps1}
\end{figure}

\clearpage
\addcontentsline{toc}{section}{Index}
\flushbottom
\printindex

\end{document}